\documentclass[%
reprint,
amsmath,amssymb,
aps,
]{revtex4-1}
\usepackage{tikz}
\usepackage{graphicx}% Include figure files
\usepackage{dcolumn}% Align table columns on decimal point
\usepackage{bm}
\usepackage{mathtools}% bold math
\usepackage{multirow}
\usepackage{color}

\newcommand{\D}{\mathrm{d}}
\newcommand{\e}{e}

\newcommand{\ex}[1]{\left<  #1 \right>}
\newcommand{\ext}[1]{\left<  #1 \right>}

\newcommand{\exqst}[1]{\left<  #1 \right>_{\mathrm{GB}}}
\newcommand{\exeq}[1]{\left<  #1 \right>_{\mathrm{B}}}

\newcommand{\kT}{\mathrm{k_{\mathrm{B}}}T}

\newcommand{\Pl}[1]{P_{\mathrm{GB}}(#1)}

\begin{document}

	\title{A work fluctuation theorem for a Brownian particle in a non confining potential}% Force line breaks 
	
	\author{Christoph Strei\ss nig, Holger Kantz }
	%\author{}%
	
	\affiliation{
		Max Planck Institute for the Physics of Complex Systems\\
		N\"othnitzer Stra\ss e 38
		D-01187 Dresden  
	}

	\begin{abstract}
		Using the Feynman-Kac formula, a work fluctuation theorem for a Brownian particle in a non confining potential, e.g., a potential well with finite depth, is derived. The theorem yields an inequality that puts a lower bound on the average work needed to change the potential in time. In comparison to the Jarzynski equality, which holds for confining potentials, an additional term describing a form of energy related to the never ending diffusive expansion appears. 	
	\end{abstract}

	\maketitle

	\section{\label{sec:1}Introduction}
	Thermal equilibrium is one of the most fundamental concepts in statistical mechanics. Roughly speaking it is a state where time does no longer appear in any of the relevant macroscopic observables. These equilibrium states are very well studied and there exists a set of basic statistical and thermodynamic statements about them.
	Let us list some of them for the simple special case of an overdamped one dimensional Brownian particle in thermal equilibrium with a heat bath of temperature $T$ and inside a potential $V(x)$. Equilibrium statistical mechanics tells us that the probability density function (PDF) of the particles position is Boltzmann distributed and hence given by 
	\begin{align}
		P_{\mathrm{B}}(x)=\frac{\e^{-V(x)\beta}}{Z_V}. \label{Boltzmann_PDF}
	\end{align}
	Here $\beta = 1/\kT$ and $Z_V$ is a normalization factor. From equilibrium thermodynamics we know that an isothermal and quasi-static transition from one equilibrium state to another, which in our case is done by changing the potential from $V_1$ to $V_2$, consumes an average amount of Energy in the form of work $W$  given by
	\begin{align}
	\ex{W} = \Delta F \label{2ndlaw_Eq}.
	\end{align}
	Where $\Delta F$ is the Helmholtz free energy difference  between the initial and the final state. 
	 Recall that $\Delta F$  is connected to the normalization factor via $\Delta F=-\mathrm{k}_{\mathrm{B}}\left[\log(Z_{V_2})-\log(Z_{V_1})\right]$.
%	\begin{align}
%		\Delta F=-\frac{1}{\beta}\ln \left(\frac{Z_{V_2}}{Z_{V_1}}\right).
%	\end{align}
	Also note, due to the stochastic nature of the system $W$ is a random variable and $\ex{\cdots}$ denotes the expectation value.
	Relaxing the quasi-static assumption the above equality (\ref{2ndlaw_Eq}) becomes an inequality
	\begin{align}
	\ex{W} \geqslant \Delta F \label{2ndlaw_in},
	\end{align}
	which can be derived by applying the Clausius inequality, a manifestation of the second law of thermodynamics, to the first law of thermodynamics.
	Surprisingly the above inequality can also be derived by a more fundamental equality, namely the Jarzynski equality  \cite{jarzynski1997nonequilibrium}
	\begin{align}
	\ex{e^{-\beta( W-\Delta F)}} = 1. \label{jar}
	\end{align} 
	This equality belongs to a family of so-called integral fluctuation theorems. In the past years a number  of integral and so-called detailed fluctuation theorems for different cases have been discovered, see \cite{seifert2012stochastic, seifert2005entropy, sagawa2010generalized, hatano2001steady,crooks1999entropy, esposito2010three,dabelow2019irreversibility, jarzynski2011equalities,neri2020second,neri2019integral} for further reading. 
	Since changing the potential with nonzero speed drives the system away from equilibrium,
	inequality (\ref{2ndlaw_in}) and the Jarzynski equality (\ref{jar}) are actually out of equilibrium results.
	Hence it is only required that the system starts in equilibrium and the final equilibrium state exists.
	The emphasis here is on exists,  $W$  does not care if the system relaxes back to equilibirum after the potential has been changed. Now for some systems equilibrium states do not exist. For our simple case, thermal equilibrium can be reached under the condition that the system is enclosed by a potential which diverges faster than logarithmically in space, e.g., a harmonic potential or hard reflective walls. We will call such potentials confining. In principle,  since most of the fundamental forces (weak, electromagnetic, gravity)
	are not diverging it is natural to assume that in reality confining potentials are very exotic. In most cases they are only local approximations 
	of globally non confining potentials, for example a harmonic 
	potential can approximate the Lennard-Jones potential around its minimum.
	
	The general question that this article is trying to tackle is the following:
	Do thermodynamic equalities and inequalities, structurally similar to the Jarzynski equality(\ref{jar}), and the lower bound (\ref{2ndlaw_in}), also exist in non-confined systems? Or in other words, how important is it to confine the system in order to get these fundamental results?
	To seek for a general answer is most certainly too ambitious, hence we constrain ourselves to the  special case of a Brownian particle inside an asymptotically flat potential which goes to zero at least as fast as $1/x$ and is changed in time via an external protocol.
	This choice is mainly motivated by the following already existing results. It was shown in \cite{aghion2019non,aghion2020infinite}  
	that for these kind of systems, assuming that the potential is time independent, to leading-order in the long time limit, the PDF  $P(x,t)$ assumes the shape  
	\begin{align}
	P(x,t)\approx \Pl{x,t}  = \frac{\e ^{ -\frac{x^2}{4Dt} - \beta V(x)}}{N(t)}, 
	\label{GaussianBoltzmannPDF}
	\end{align} 
	where $N(t)$ is the normalization constant which is  $\sim \sqrt{t}$ for sufficiently large $t$.
	Eq. \eqref{GaussianBoltzmannPDF}, has a simple intuitive explanation: The Gaussian factor in the asymptotic shape of the PDF is dominant in the tails of the system, at $x>\sqrt{\pi Dt}$ where the potential is effectively zero whereas at small $x$ and $t\gg1$, the Gaussian factor is $=1$ and the Boltzmann factor is dominant. When $t\rightarrow\infty$, according to Eq. \eqref{GaussianBoltzmannPDF}, the PDF approaches a non-normalizable Boltzmann infinite invariant density \cite{aghion2019non,aghion2020infinite} (see also the related works   \cite{dechant2011solution,wang2019ergodic}) $\lim_{t\rightarrow\infty} N(t)P_t(x)\rightarrow \exp(-V(x)/\mathrm{k}_{\mathrm{B}} T)$, which replaces the standard Boltzmann distribution  in its role in  determining integrable physical observables such as  energy and occupation times, and leads to infinite ergodic theory, see e.g.,  \cite{aaronson1997introduction,dechant2011solution,leibovich2019infinite, meyer2017infinite,akimoto2020infinite}.
	
	\section{\label{sec:2}Setting the stage}
	
	We begin with the overdamped Langevin dynamics of a Brownian particle in an external potential field
	\begin{align}
	\dot{x}_t = -\frac{V'(x_t,\lambda_t)}{\gamma} + \sqrt{2 D}\; \xi_t, \label{Langevin}
	\end{align}
	where $V(x_t,\lambda_t) $ is a potential depending on an externally controlled protocol $\lambda_t$, and $D$, $\gamma$, $\xi_t$ are  respectively the diffusion constant, the friction and Gaussian white noise with zero mean and 
	\begin{align}
	\ex{\xi_{t}\,\xi_{t'}} = \delta(t-t').
	\end{align}
	Furthermore $V(x_t,\lambda_t)$ is assumed to be an asymptotically flat potential well which falls of at least as rapidly as $1/x$ hence 
	\begin{align}
	\lim\limits_{x \rightarrow \pm \infty}V(x,\lambda_t) = 0.
	\end{align}
	The evolution  of the PDF $P(x,t)$ is given by 
	\begin{align}
	\partial_t P(x,t) = L  P(x,t) ,
	\end{align}
	where $L$ is the Fokker-Planck operator
	\begin{align}
	L = \left(D\partial_x^2 + \frac{1}{\gamma}\partial_x V'\right).
	\end{align}
	For a fixed $\lambda_t$ and sufficiently long times $P(x,t)$  converges to \cite{aghion2020infinite}
	\begin{align}
	\Pl{x,t,\lambda_t}  = \frac{\e ^{ -\frac{x^2}{4Dt} - \beta V(x,\lambda_t)}}{N(t,\lambda_t)}, \label{uniformdens}
	\end{align} 
	here $\beta = \frac{1}{\kT}$, $\mathrm{k_B}$ is the Boltzmann constant
	and $N(t,\lambda_t)$ is the normalization constant
	\begin{align}
	N(t,\lambda_t) = \int_{-\infty}^{\infty} \e ^{ -\frac{x^2}{4Dt} - \beta V(x,\lambda_t)} \D x.
	\end{align}
	Although we mentioned in the introduction that for large enough $t$, $N(t,\lambda_t) \sim \sqrt{t}$ , we choose to keep the 
	full normalization constant since it leads to faster convergence.
	
	The particular scenario that we consider throughout this article is the following.
	At $t=0$ the particle is placed inside the potential well. From $t=0$ to $t=t_0$ the system relaxes such that at $t=t_0$ the density is approximately given by $\Pl{x,t_0}$, Eq. \eqref{uniformdens}.
	From $t=t_0$ to $t=t_1$ the potential is changing according to an externally controlled protocol $\lambda_t$. At $t=t_1$ the potential stops changing and in principle the system relaxes back to a state described by (\ref{uniformdens}). The relaxation in the end however will not play a role in the results.
	In this scenario the work done by the protocol along a trajectory up to time $t$ is given by 
	\begin{align}
	W_t &=  \int_{t_0}^{t}\dot{\lambda}_{\tau}\frac{\partial V(x_{\tau},\lambda_{\tau})}{\partial \lambda_{\tau}} \D \tau = \int_{t_0}^{t} \frac{\partial V(x_{\tau},\tau)}{\partial \tau} \D \tau .\label{workdef}
	\end{align}
	
	\section{A motivating special case: The infinitely fast protocol \label{sec:3A}}
	Let us start by considering a simple special case where the potential changes instantaneously.
	This can be expressed mathematically by stating that the change of the potential $V(x,\Theta(t-t_0))$ in time is only through a heaviside/theta function $\Theta(t-t_0)$.
	The natural choice for the protocol here is 
	\begin{align}
	\lambda_t =\Theta(t-t_0).   
	\end{align}
	Introducing the abbreviate notation
	\begin{align}
	\Delta V(x)\coloneqq V(x,1) - V(x,0)
	\end{align}
	we write the potential as
	\begin{align}
	V(x,\lambda_t) = V(x,0) + \lambda_t \Delta V(x).
	\end{align}
	According to \eqref{workdef} the trajectory dependent work is then given by the difference between the potential after and before the change evaluated at $x_{t_0}$,
	\begin{align}
	W_t =  \Delta V(x_{t_0}). 
	\end{align} 
	As mentioned in the introduction we are interested in a Jarzynski like equality. Due to the simple expression for the work  we can straight forwardly calculate
	\begin{align}
	\ex{\e ^{-\beta W_t }} &=  \int_{-\infty}^{\infty} \e ^ {- \Delta V(x)} \; \Pl{x,t_0,0} \; \D x \\  
	&=  \frac{\int_{-\infty}^{\infty} \e ^{ -\frac{x^2}{4Dt_0} - \beta V(x,1)}\D x}{N(t_0,0)}. \label{rhs_jar}					 						  
	\end{align}
	 Introducing a quantity $ \Delta G$ analogue to the Helmholtz free energy difference
	\begin{align}
	\Delta G &=-\beta \ln\left(\frac{N(t_0,0)}{N(t_0,1)}\right)\label{fastG}\\ &=-\beta \ln\left(\frac{\int_{-\infty}^{\infty} \e ^{ -\frac{x^2}{4Dt_0} - \beta V(x,1)}\D x}{\int_{-\infty}^{\infty} \e ^{ -\frac{x^2}{4Dt_0} - \beta V(x,0)}\D x}\right)
	\end{align}
	we arrive at
	\begin{align}
	\ex{e^{-\beta W_t}} =  \e^{-\beta \Delta G}  \label{jar_ana}.
	\end{align}
	Eq. \eqref{jar_ana} is analogous to Eq. \eqref{jar}, but in contrast to the standard Jarzinski equality, it is now valid even though the system has no equilibrium state. By the so called Jensen's inequality this relation yields
	\begin{align}
	\ex{W_t} \geqslant  \Delta G .\label{flucttheo}
	\end{align}
	In the next section we will derive a version of \eqref{jar_ana} valid 
	for arbitrary protocol speed. 

	\section{Derivation of the work fluctuation theorem \label{sect3B}}
	Our derivation is essentially an adjusted version of an elegant derivation of the Jarzynski equality using the Feynman-Kac formula, first presented in \cite{hummer2001free}.
	Let us briefly state a version of the Feynman-Kac formula which is sufficient for our purpose, for a proof see \cite{boksenbojm2010nonequilibrium}. 
	Assume a Langevin process $x_t$ whose phase space density 
	$P(x,t) = \ext{\delta(x_t-x)}$ obeys 
	\begin{align}
	\partial_t P(x,t) = L  P(x,t) .
	\end{align}
	Here $\ext{\cdots}$ denotes an average over all trajectories ending at time $t$ and $\delta(x_t-x)$ being the delta-distribution picks out the ones that 
	end at position $x$.
	The Feynman Kac formula then says that 
	\begin{align}
	g(x,t) = \ext{\delta(x-x_t) \e^{-\Omega_t}}, \label{FKsol}
	\end{align}
	with 
	\begin{align}
	\Omega_t = \int_{t_0}^{t} f(x_{\tau},t) \D \tau,
	\end{align}
	being a stochastic functional obeys 
	\begin{align}
	\partial_t g(x,t) = L  g(x,t) - f(x,t) g(x,t) .\label{FK}
	\end{align}
	Now we apply this statement to our case
	by making the initially arbitrary seeming choice 
	\begin{align}
	\Omega_t \coloneqq\beta\left[ W_t  -  \int_{t_0}^{t} \left(\frac{\kT}{2 \tau }  + \frac{x_{\tau}F(x_{\tau},\lambda_{\tau})}{2 \tau}\right) \D \tau\right], \label{omegachoice}
	\end{align}
	or equivalently 
	\begin{align}
	f(x,\tau) \coloneqq \beta  \left[\frac{\partial V(x,\tau,\lambda_{\tau})}{\partial \tau}  -   \frac{\kT}{2 \tau }  - \frac{x_{\tau}F(x_{\tau},\lambda_{\tau})}{2 \tau}\right], 
	\end{align}
	with $F=-V'$ being the force acting on the particle.
	Equation $(\ref{FK})$ then becomes
	\begin{align}
	\partial_t g(x,t)\nonumber &=  L  g(x,t)\\&+\beta\left[\frac{\kT}{2 t}  +  \frac{x F(x,t)}{2 t}- \dot{\lambda}\frac{\partial V(x,\lambda_t)}{\partial \lambda_t}\right] g(x,t). \label{FKS}
	\end{align}
	It can be verified by direct substitution that 
	\begin{align}
	g(x,t)= \frac{\e ^{ -\frac{x^2}{4Dt} - \beta V(x,\lambda_t)}}{N(t_0,\lambda_{t_0})},
	\end{align}
	solves  (\ref{FKS}) with the initial condition 
	\begin{align}
	g(x,t_0) \equiv P(x,t_0) =\Pl{x,t_0,\lambda_{t_0}}.
	\end{align}
	However we also know from the Feynman-Kac formula that (\ref{FKsol}) with the particular choice made in (\ref{omegachoice}) solves 
	(\ref{FKS}). Thus we have 
	\begin{align}
	\ext{\delta(x-x_t) \e^{-\Omega_t}}= \frac{\e ^{ -\frac{x^2}{4Dt} - \beta V(x,\lambda_t)}}{N(t_0,\lambda_{t_0})},
	\end{align}
	which can be rewritten by defining a more general analogue of the Helmholtz free energy difference than \eqref{fastG} 
	\begin{align}
	\Delta G \coloneqq -\kT \ln \left( \frac{N(t,\lambda_t)}{N(t_0,\lambda_{t_0})} \right),
	\end{align}
	as
	\begin{align}
	\ext{\delta(x-x_t) \e^{-(\Omega_t-\beta \Delta G)}}= \Pl{x,t,\lambda_t}.\label{detailed}
	\end{align}
	Integration over $x$ and using (\ref{omegachoice}) gives a work integral fluctuation theorem
	\begin{align}
	\ex{ \e^{-\beta\left[ W_t  -  \int_{t_0}^{t} \left(\frac{\kT}{2 \tau }  + \frac{x_{\tau}F(x_{\tau},\lambda_{\tau})}{2 \tau}\right) \D \tau-\Delta G\right]}}= 1. \label{work_int}
	\end{align}
	which by applying the Jensen's inequality yields 
	\begin{align}
	\ex{W_t}  \geqslant \Delta G + \ex{\int_{t_0}^{t} \left(\frac{\kT}{2 \tau }  + \frac{x_{\tau}F(x_{\tau},\lambda_{\tau})}{2 \tau}\right)\D \tau}. \label{in}
	\end{align}
	The Fluctuation theorem given by Eq.(\ref{work_int}) is the central result of this article. Its physical meaning will be discussed in the next section.  
	
	\section{A possible physical Interpretation}
	
	Let us investigate the terms appearing in the exponent of the fluctuation theorem  Eq.(\ref{work_int}) in more detail. One major difference with respect to the Jarzynski equality is the additional trajectory dependent term
	\begin{align}
	\int_{t_0}^{t} \left(\frac{\kT}{2 \tau }  + \frac{x_{\tau}F(x_{\tau},\lambda_{\tau})}{2 \tau}\right)\D \tau \label{new_term}.
	\end{align} 
	
	Another minor difference is that the time dependence of $\Delta G$ is not only  due to the protocol but also explicitly due to the Gaussian term in the normalization constant. It is clear that both of these discrepancies are a mathematical consequence of the non-equilibrium initial PDF. Using the Feynman-Kac derivation scheme, as presented in the previous section, one could in principle derive 
	an integral fluctuation theorem similar to (\ref{work_int}) for any kind of non-equilibrium initial PDF. However,  $\Pl{x,t_0,\lambda_{t_0}}$ being the long time asymptotic density lets us expect that for a sufficiently slow protocol i.e. in the quasi-static limit  $P(x,\tau,\lambda_{\tau}) = \Pl{x,\tau,\lambda_{\tau}}$ for $\tau \geqslant t_0$. We will support this claim later with numerical evidence. Let us now calculate 
	\begin{align}
	&\exqst{\Omega_t} = \int_{t_0}^{t}\D 
	\tau \exqst{f(x,\tau, \lambda_{\tau})}\label{first_line}\\& = \int_{t_0}^{t}\D 
	\tau \int_{-\infty}^{\infty} \D x f(x,\tau,\lambda_{\tau}) \frac{\e^{-\frac{x^2}{4D\tau}- \beta V(x, \lambda_{\tau})}}{N(\tau, \lambda_{\tau})}\label{second_line} \\ &=- \int_{t_0}^{t}\D 
	\tau \frac{1}{N(\tau, \lambda_{\tau})} \int_{-\infty}^{\infty} \D x (\partial_t - L)\e^{-\frac{x^2}{4D\tau}-\beta V(x, \lambda_{\tau})} \label{third_line} \\ & = \Delta G.
	\end{align}
	Here $\exqst{\cdots}$ denotes the expectation value with respect to $\Pl{x,\tau,\lambda_{\tau}}$
	or in other words the expectation value in the quasi-static limit. Note, from line (\ref{second_line}) to (\ref{third_line}) Eq.(\ref{FKS}) respectively Eq.(\ref{FK}) was used. 
	Writing $\Omega_t$ explicitly using Eq.(\ref{omegachoice})  we get 
	\begin{align}
	\exqst{W_t}  = \Delta G + \exqst{\int_{t_0}^{t} \left(\frac{\kT}{2 \tau }  + \frac{x_{\tau}F(x_{\tau},\lambda_{\tau})}{2 \tau}\right)\D \tau}. \label{qusteq}
	\end{align}
	The equation above shows that in the quasi-static limit inequality (\ref{in}) becomes an equality.
	The analogue statement for confined systems is that for sufficiently slow protocols the system stays Boltzmann distributed which leads to $\exeq{W_t} = \Delta F$, where $\exeq{\cdots}$ denotes the average with respect to the Boltzmann density (\ref{Boltzmann_PDF}). However there is a very intriguing difference between these two statements.
	For cyclic protocols, meaning $\lambda_{t_0} = \lambda_{t}$,
	applied to confined systems it is clear that $\exeq{W_t} = 0$ since $\Delta F = 0 $.
	Whereas for cyclic protocols applied to non confined systems it is not obvious 
	from $(\ref{qusteq})$ whether $\exqst{W_t} = 0$. 
	This raises the question if its possible to get $\exqst{W_t} \leqslant 0 $ or more generally $\ex{W_t} \leqslant 0 $.
	Or in other words is it possible to extract energy in the form of work by applying a cyclic protocol? It is important to realize that due to the never-ending diffusive process a cyclic protocol does not mean that the system itself returns to its initial state. For now we will leave this question open and approach it numerically in the next section.
	So far we can make the following conclusions:
	\begin{align}
	\Delta G + \ex{\int_{t_0}^{t} \left(\frac{\kT}{2 \tau }  + \frac{x_{\tau}F(x_{\tau},\lambda_{\tau})}{2 \tau}\right)\D \tau} \label{quantitiy}
	\end{align}
	is a quantity that puts a lower bound on the average work needed to externally change the potential. In the quasi-static limit this quantity becomes
	the average work and if negative it is free to use for the external observer.
	It should be mentioned that due to the protocol dependence of second term it is not something like a free energy in the sense of an thermodynamic potential like the Helmholtz free energy. 
	
	Let us now focus on the second term in (\ref{quantitiy}). Since it originates from the Gaussian part of $\Pl{x,\tau,\lambda_{\tau}}$ we claim, at least in the quasi-static limit,
	that it can be interpreted as an energy coming from the expansion of the system.
	And indeed it can be brought into a convenient form resembling pressure-volume work. In order to do that we first need to establish a notion of pressure.
	The osmotic pressure $\Pi$ of a Brownian particle confined in a region of size $L$ and inside a force field $F(x)$ is given by \cite{brady1993brownian}
	\begin{align}
	\Pi = \frac{1}{L} \left[\kT + \ex{x F(x) } \right]. \label{osmotic}
	\end{align}	
	Of course our system is not confined so it is questionable how to make use of the  above expression, especially how to choose  the size of the system. Nevertheless, choosing the 
	length scale of diffusion  $L_{\tau} = \sqrt{2D\tau}$ as a  measure for the size of the system and introducing a quantity  
	\begin{align}
	p_{\tau} \coloneqq \frac{1}{L_{\tau}} \left[\kT + F(x_{\tau}, \lambda_{\tau}) x_{\tau}\right], \label{pressure}
	\end{align}
	which can be seen as an analogue of $\Pi$ but for a single particle,
	 allows us to rewrite 
	\begin{align}
		\int_{t_0}^{t} \left(\frac{\kT}{2 \tau }  + \frac{x_{\tau}F(x_{\tau},\lambda_{\tau})}{2 \tau}\right)\D \tau = \int_{L_{t_0}}^{L_t} p_{\tau} \,  \D L_{\tau}.
	\end{align}
	Here we have substituted $\tau = L_{\tau}^2/(2D)$ in the integral and used definition \eqref{pressure}.
	Consequently Eq.(\ref{work_int}) and inequality (\ref{in}) can be written as
	\begin{align}
	\ex{ \e^{-\beta\left[ W_t  -  \int_{L_{t_0}}^{L_t} p_{\tau} \,  \D L_{\tau}-\Delta G\right]}}= 1, \label{FT_pressure}
	\end{align} 
	and 
	\begin{align}
	\ex{W_t}  \geqslant \Delta G + \ex{\int_{L_{t_0}}^{L_t} p_{\tau} \,  \D L_{\tau}}. 
	\end{align} 	
	We agree that the structure of the integral in $(\ref{FT_pressure})$ could just be a nice coincidence. However let us present another argument.
	Assume a one dimensional Brownian particle with diffusion coefficient $\tilde{D}$ and temperature $\tilde{T}$ inside a confining potential $\tilde{V}(x,t)$ given by 
	\begin{align}
	\tilde{V}(x,\tau) = V(x,\tau) + \frac{x^2}{4D\tau} \kT.
	\end{align}
	Note  $V(x,\tau)$ is as before a non confining potential but $\tilde{V}(x,\tau)$ is now enclosed by an additional harmonic potential which opens up with time. 
	In the quasi-static limit the PDF of the system is given by $\Pl{x,\tau}$ and is thus indistinguishable from our non-confined system.
	The average work in the confined system yields 
	\begin{align}
	&\exqst{\tilde{W}_t}= \exqst{ \int_{t_0}^{t}\frac{\partial \tilde{V}(x_{\tau}, \tau)}{\partial \tau} \D \tau} \label{conline1} \\
	&=
	\exqst{ \int_{t_0}^{t}\partial_{\tau} V(x_{\tau}, \tau)\D \tau}-\kT\exqst{ \int_{t_0}^{t} \frac{x_{\tau}^2}{4D\tau^2} \D \tau} \nonumber \\
	&= \int_{t_0}^{t} \D \tau \exqst{ \partial_{\tau}V(x,\tau) - \kT \frac{x^2}{4D\tau^2}} \\
	&=-\int_{t_0}^{t} \D \tau \frac{1}{N(\tau)} \int \D x \,  \partial_{\tau} \left(\e ^{ -\frac{x^2}{4Dt} - \beta V(x,\tau)}\right) \label{conline3} \\
	& = \exqst{W_t} - \exqst{\int_{t_0}^{t} \left(\frac{\kT}{2 \tau }  + \frac{x_{\tau}F(x_{\tau},\tau)}{2 \tau}\right)\D \tau} \label{conline4}\\
	& = \exqst{W_t} -\exqst{\int_{L_{t_0}}^{L_t} p_{\tau} \,  \D L_{\tau}} \label{conline5},
	\end{align}
	where Eq. \eqref{FKS} was used to get from line \eqref{conline3} to line \eqref{conline4} , note the vanishing boundary terms.
	The above calculation shows that in the quasi-static limit the work done by opening the harmonic potential coincides with the path dependent part of the expansion energy of the non-confined system. The main difference between the two forms of energy is that in the confined system work is assumed to be externally controllable. 
	In the non-confined system a part of the explicit time dependence comes from the inherent diffusive expansion and is hence not assumed to be externally controllable.
    \section{Examples \label{sec:Numerical Verification}}
\begin{figure}
	\includegraphics[width = 8 cm]{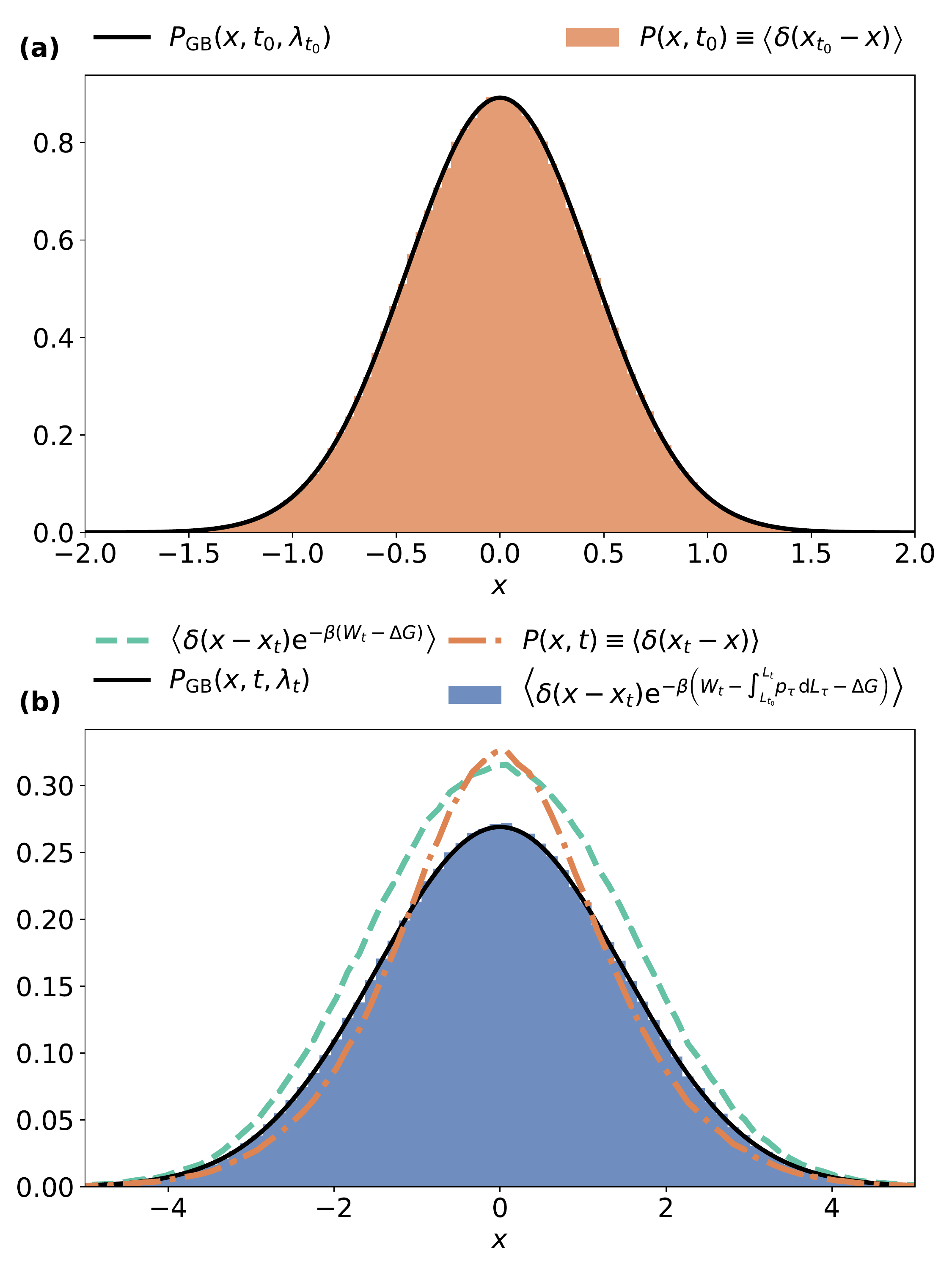}
	\caption{\footnotesize Numerical results for  Brownian particle inside a potential given by (\ref{potential}), parameters are chosen a s follows: $ D = \kT = \gamma = 1$, $A(\tau) =  \theta(\tau - t_0) \sin(\frac{\tau- t_0}{t-t_0}\pi)$, $B(\tau) = 1$,  $ t_0 = 0.1$, $t = 1$, $\Delta \tau = 10^-3$, $n=10^6$. (a) Shows a comparison at time $\tau=t_0$ of the analytic expression \eqref{uniformdens} for asymptotic long time density $\Pl{x,t_0,\lambda_{t_0}}$ (black solid line) with a histogram (orange filled histogram) constructed from an ensemble of numerically generated trajectories representing the PDF $P(x,t_0)$.  (b) Shows a comparison at $\tau = t$ of the analytic expression \eqref{uniformdens} for the asymptotic long time density $\Pl{x,t,\lambda_{t}}$(black solid line) with three different histograms.
	Each of these histograms is constructed from an ensemble of numerically generated trajectories. The dash dotted orange line represents the regular PDF $P(x,t)$. 
	The blue filled histogram represents the left-hand side of Eq. \eqref{detailed},
	the path probabilities are thus re-scaled by  $\exp\left({-\beta\left[ W_t  -  \int_{L_{t_0}}^{L_t} p_{\tau} \,  \D L_{\tau}-\Delta G\right]}\right)$.
	The green dashed line represents a histogram where path probabilities are re-scaled by $\exp\left({-\beta\left[ W_t -\Delta G\right]}\right)$ }, emphasizing the relevance of the "pressure-volume" term".\label{fig:sin}
\end{figure}
\begin{figure}
	\begin{center}
		\includegraphics[width = 8 cm]{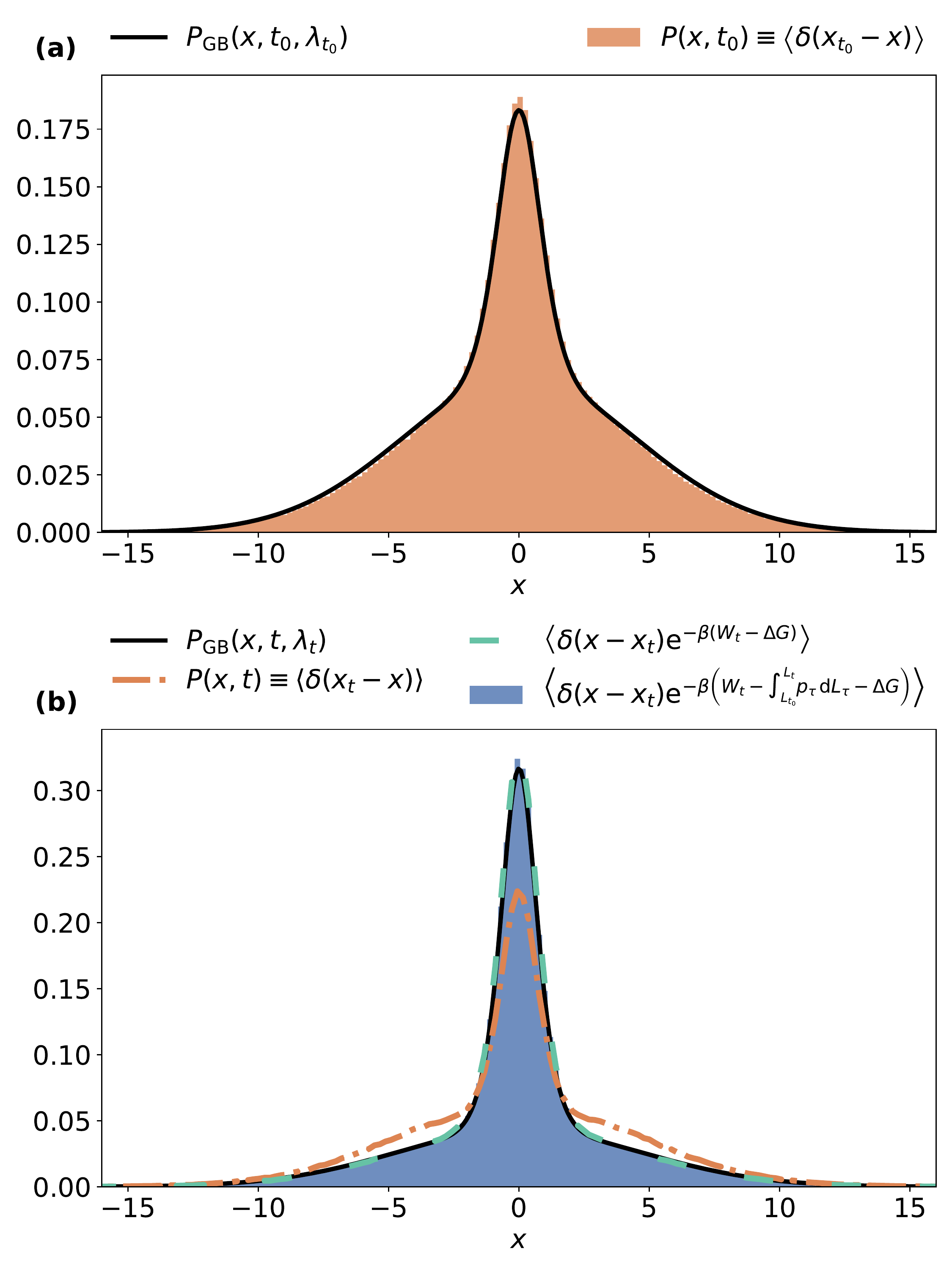}
		\caption{\footnotesize
			Same scenario as described in Fig.\ref{fig:sin} with parameters chosen as follows:
			$ D = \kT = \gamma = 1$, $A(\tau) =  \theta(\tau - t_0)\frac{\tau- t_0}{t-t_0}+1$, $B(\tau) = 1$,  $ t_0 = 10$, $t = 11$, $\Delta \tau = 10^-3$, $n=10^6$. 
		}
		\label{fig:amp}
	\end{center}
\end{figure}
\begin{figure}
	\begin{center}
		\includegraphics[width = 8 cm]{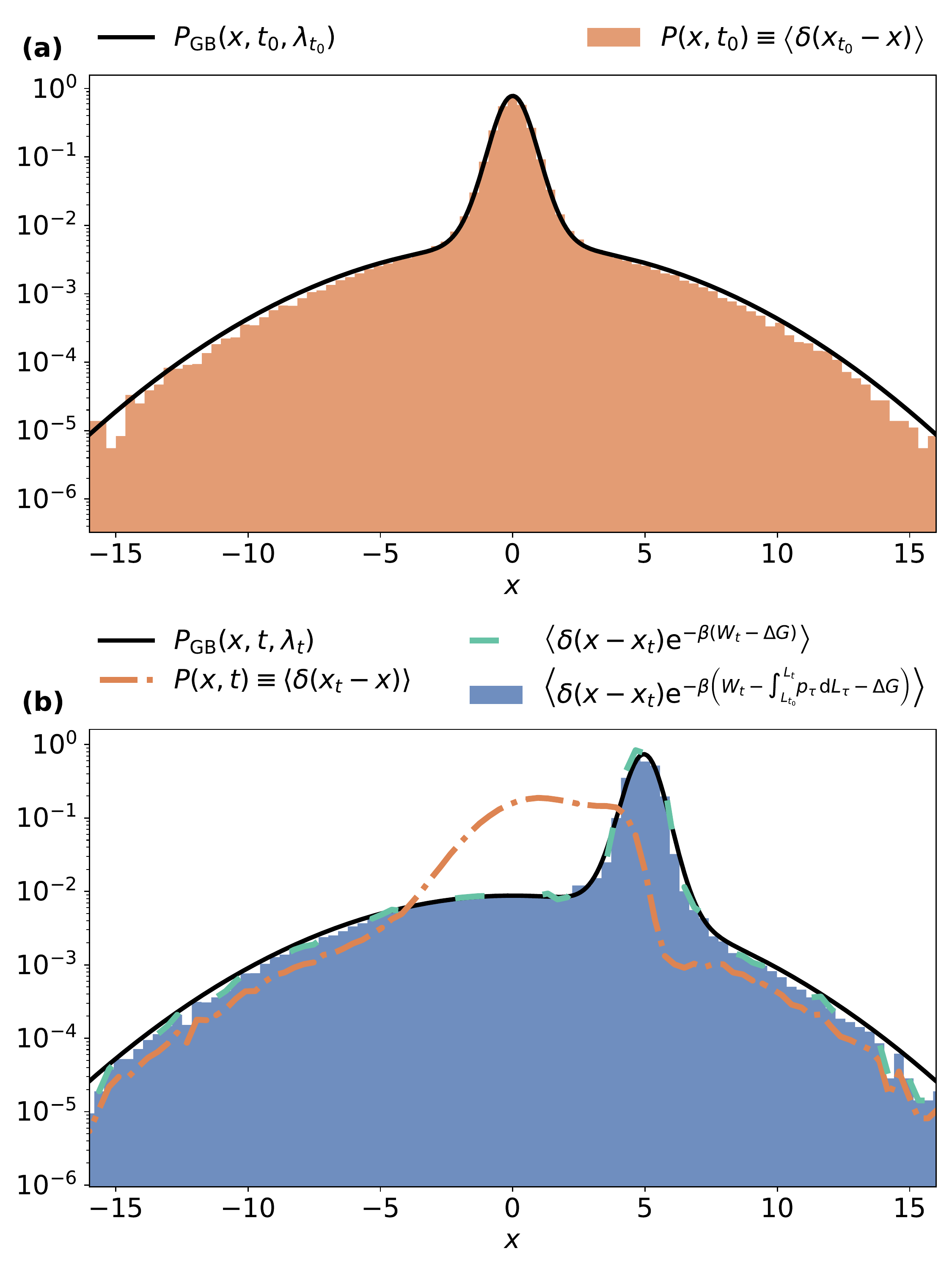}
		\caption{\footnotesize
			Same scenario as described in Fig.\ref{fig:sin} with parameters chosen as follows:
			$ D = \kT = \gamma = 1$, $A(\tau) = 5$, $B(\tau) = 5\theta(\tau - t_0) \frac{\tau- t_0}{t-t_0} $,  $ t_0 = 10$, $t = 11$, $\Delta \tau = 10^-3$, $n=10^6$. Note the logarithmically scaled y axis.
		}
		\label{fig:mov}
	\end{center}
\end{figure}
\begin{figure}
	\begin{center}
		\includegraphics[width = 8 cm]{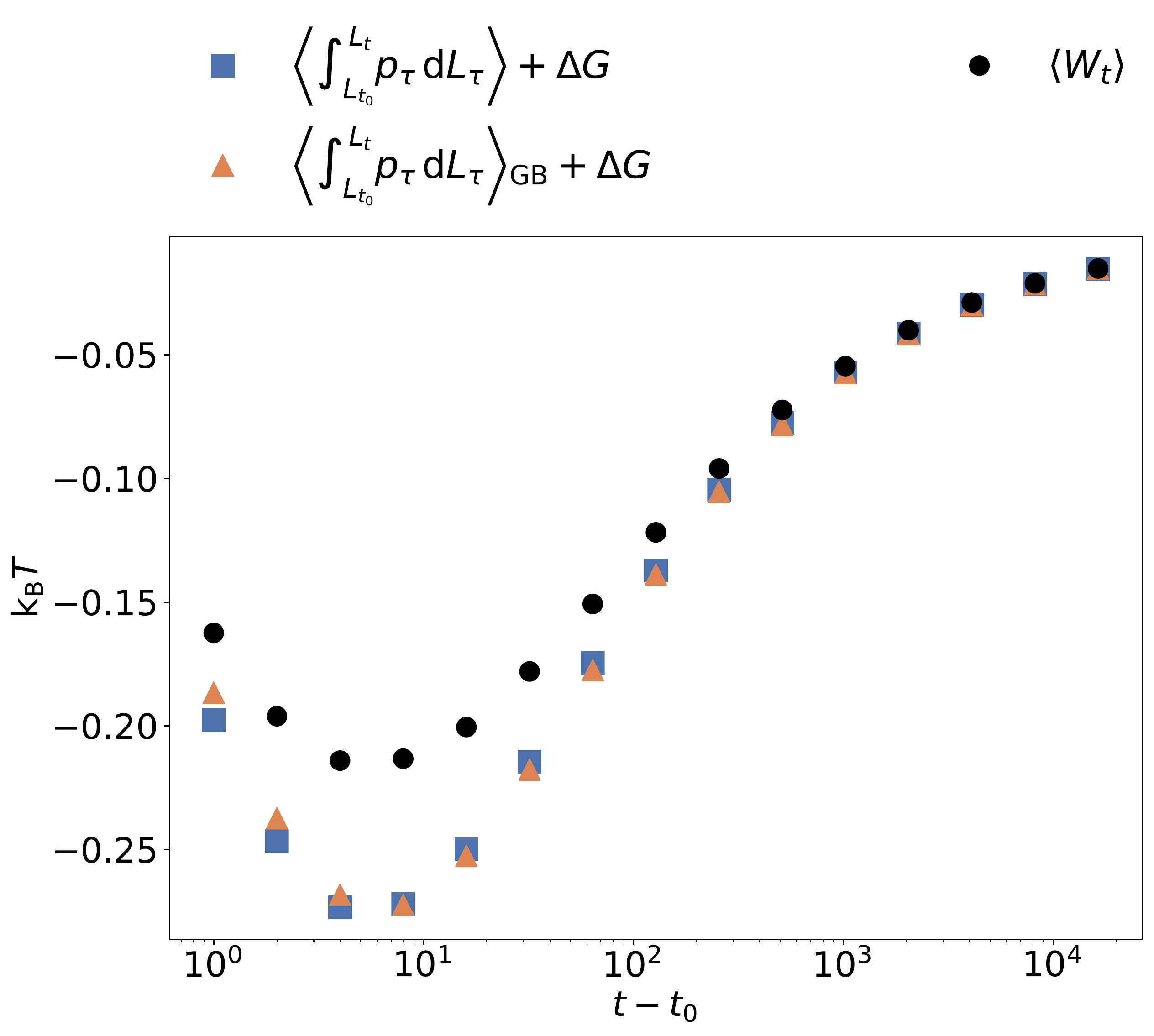}
		\caption{\footnotesize Average work (black dots) and expression (\ref{quantitiy}) (blue squares) vs. duration of the protoccol $t-t_0$. The orange triangles are displaying the semi-analytical calculation of the right-hand-side of Eq. \ref{qusteq}.
			Parameters are choosen as follows: $ D = \kT = \gamma = 1$, $A(\tau) =  \theta(\tau - t_0) \sin(\frac{\tau- t_0}{t-t_0}\pi)$, $B(\tau) = 1$,  $ t_0 = 0.5$, $\Delta \tau = 10^{-3}$,  $n=10^5$.
		}
		\label{fig:quasistat_sin}
	\end{center}
\end{figure}
\begin{figure}
	\begin{center}
		\includegraphics[width = 8 cm]{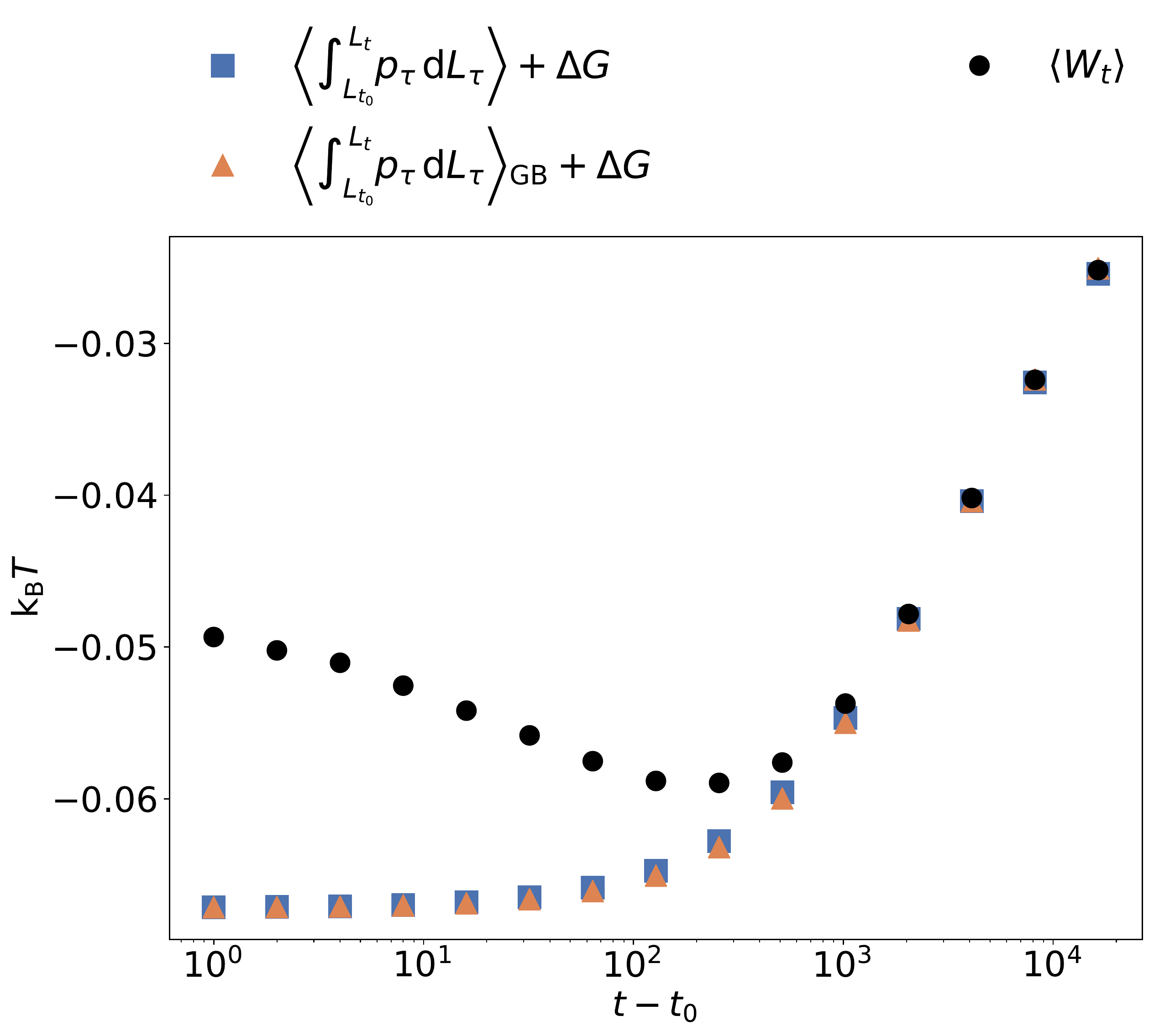}
		\caption{\footnotesize Average work (black dots) and expression (\ref{quantitiy}) (blue squares) vs. duration of the protocol $t-t_0$. The orange triangles are displaying the right-hand-side of Eq. \ref{qusteq}.
			Parameters are chosen as follows: $ D = \kT = \gamma = 1$, $A(\tau) =  \theta(\tau - t_0)\frac{\tau- t_0}{t-t_0}+1$, $B(\tau) = 1$,  $ t_0 = 10^3$, $\Delta \tau = 10^{-3}$, n = $10^5$ 
		}
		\label{fig:quasistat_amp}
	\end{center}
\end{figure}
\begin{figure}
	\begin{center}
		\includegraphics[width = 8 cm]{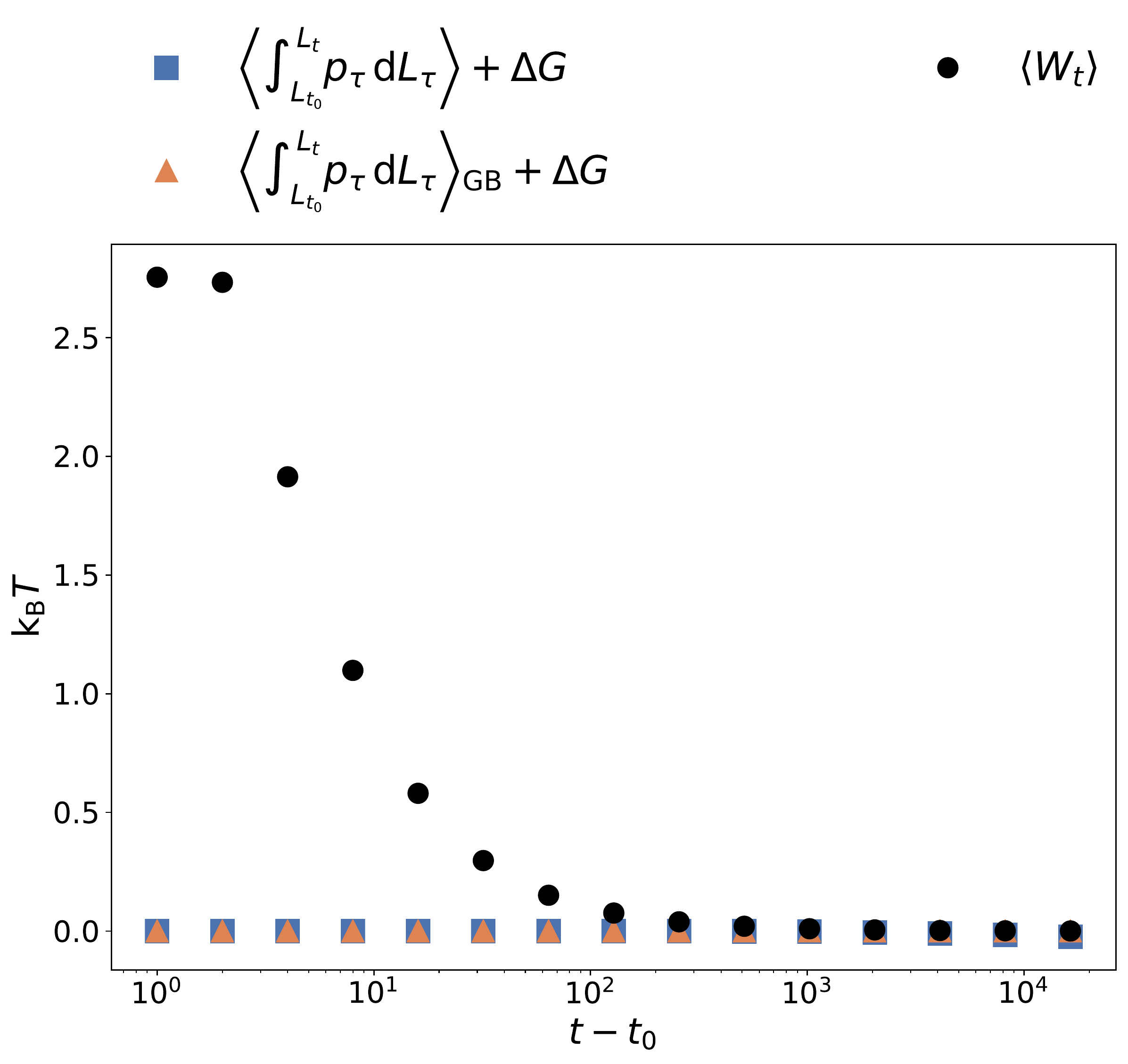}
		\caption{\footnotesize Average work (black dots) and expression (\ref{quantitiy}) (blue squares) vs. duration of the protocol $t-t_0$. The orange triangles are displaying the right-hand-side of Eq. \ref{qusteq}.
			Parameters are chosen as follows: $ D = \kT = \gamma = 1$, $A(\tau) = 5$, $B(\tau) = 5\theta(\tau - t_0) \frac{\tau- t_0}{t-t_0} $,  $ t_0 = 10^4$, $\Delta \tau = 10^{-3}$, $n=10^5$ 
		}
		\label{fig:quasistat_mov}
	\end{center}
\end{figure}
\begin{figure}
	\begin{center}
		\includegraphics[width = 8 cm]{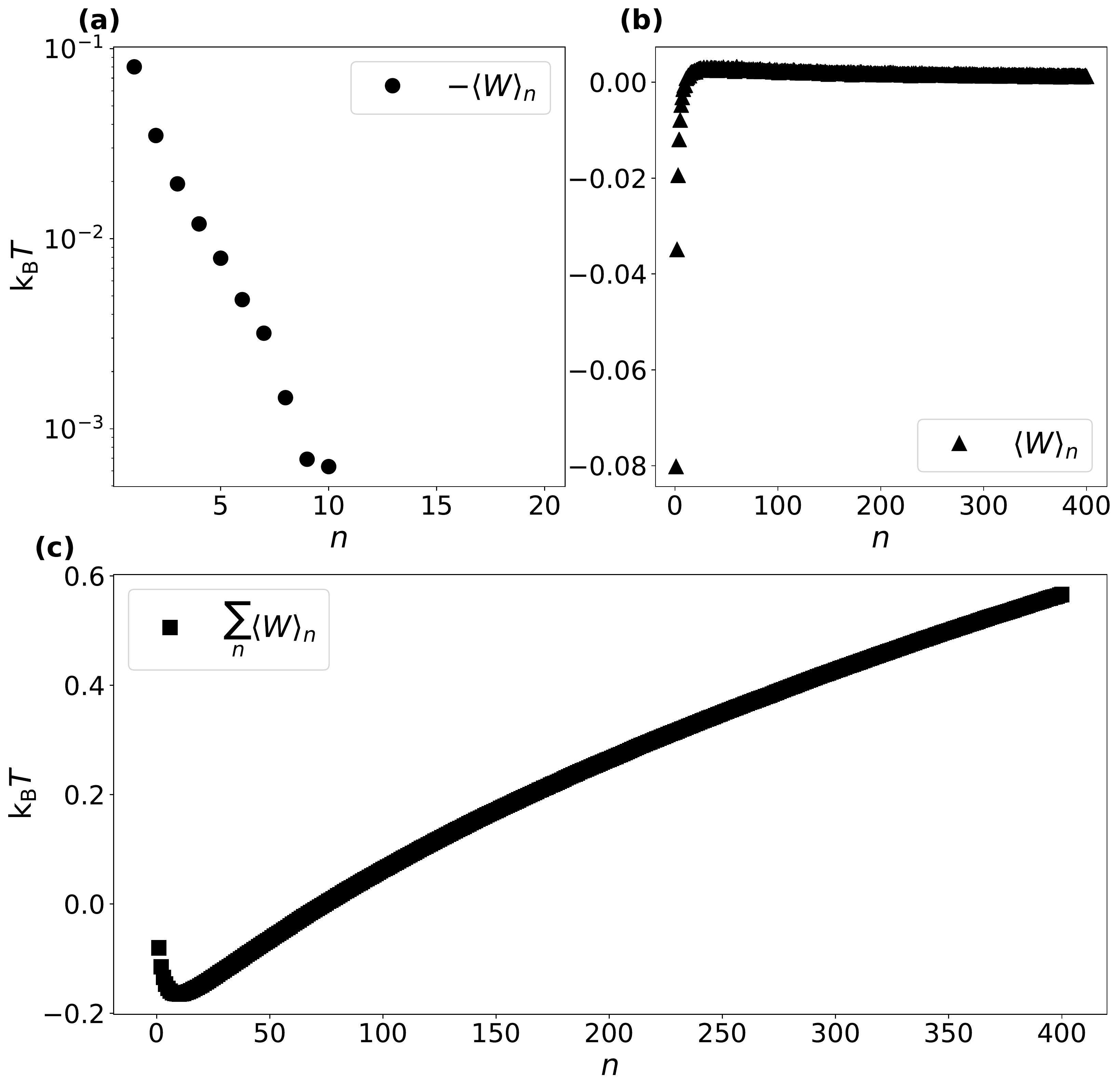}
		\caption{\footnotesize Behavior of the average work $\ex{W}_n$ per cycle $n$ for a cycle duration of 1. (a) Shows the initially exponential behavior. For $n > 9 $, $-\ex{W}_n < 0$  and hence can not be displayed in the semi-logarithmic plot.
			(b) Shows that $\ex{W}_n$ goes to a value slightly above zero.
			This results in a linear increasing cumulative sum
			$\sum_n \ex{W}_n$, as can be seen in (c) .
		}
		\label{fig:cycle}
	\end{center}
\end{figure}
As an example for our theory we choose the inverted Gaussian potential well 
\begin{align}
V(x,\lambda_{\tau})=-A(\lambda_{\tau}) \e^{-\frac{(x-B(\lambda_{\tau}))^2}{2}} \label{potential},
\end{align}
whose depth $A(\lambda_{\tau})$ or location $B(\lambda_{\tau})$ is changed in time by the 
protocol $\lambda_{\tau}$. A convenient way to show the integral fluctuation theorem  Eq.(\ref{work_int}) is by showing it indirectly via verifying  Eq.(\ref{detailed}).
We proceed in the following manner. An ensemble of $n_{\mathrm{trajectory}}$ trajectories is generated  using the standard Euler-Maruyama method with an time increment of $\Delta \tau $ and initial position $x=0$.
At $\tau = t_0$ a PDF is constructed and checked if it has converged to $\Pl{x,t_0,\lambda_{t_0}}$. At the end of the protocol which is at $\tau=t$, the PDF is checked again to make sure that it is now different 
from $\Pl{x,t,\lambda_{t}}$, which should be the case for sufficiently fast protocols. The PDF's are simply constructed as histograms from the ensemble.
To verify Eq.(\ref{detailed}) we have to recall that expectation values for stochastic processes are path integrals, namely we can write: $\langle \delta(x-x_t)\e^{-\Omega[x_{\tau}]-\beta \Delta G}\rangle=\int \mathcal{D}[x_{\tau}] \delta(x-x_t) \e^{-\Omega[x_{\tau}]-\beta \Delta G}p[x_{\tau}]$, where $p[x_{\tau}]\mathcal{D}[x_{\tau}]$ is a measure for the probability to observe a trajectory $x_{\tau}$. Plugging this into Eq. \eqref{detailed} yields 
\begin{align}
\int \mathcal{D}[x_{\tau}] \delta(x-x_t) \e^{-\left( \Omega[x_{\tau}]-\beta \Delta G\right)}p[x_{\tau}]= \frac{\e ^{ -\frac{x^2}{4Dt} - \beta V(x,\lambda_t)}}{N(t_0,\lambda_{t_0})} \label{path_FT}.
\end{align}
The form  of (\ref{path_FT}) allows us to interpret  $\e^{-\left(\Omega[x_{\tau}]-\beta \Delta G\right)}$ as an additional weight on the path-probability.
Therefore, if we multiply the increment that one particle adds to the height of a bin in the PDF's histogram by  $\e^{-\left(\Omega[x_{\tau}]-\Delta G\right)}$ we get 
an histogram representing the left hand side of Eq.(\ref{detailed}). And if Eq.\eqref{detailed} is correct this histogram should match $\Pl{x,t,\lambda_{t}}$.
The results for different cases of $A(\tau)$ and $B(\tau)$ are displayed in
Fig.(\ref{fig:sin}-\ref{fig:mov}). Let us briefly discuss them.

In Fig.\ref{fig:sin}  $A(\tau) =  \theta(t_0-\tau) \sin(\frac{\tau- t_0}{t-t_0}\pi)$ and  $B(\tau) = 1$,  which means the particle freely diffuses during the initial relaxation meaning $P(x,t_0)=\Pl{x,t_0,\lambda_{t_0}}$ is exact for an arbitrary small $t_0$ and the derivation of Eq.(\ref{detailed}) is exact as well.
As such we can see this case as a test of the numerical procedure more than a test of the analytical results.  Fig. \ref{fig:sin} (a) shows the expected agreement 
of the PDF  with $\Pl{x,t_0,\lambda_{t_0}}$.  Fig.\ref{fig:sin} (b) shows that in the end of the protocol $P(x,t,\lambda_{t})\neq\Pl{x,t,\lambda_{t}}$ furthermore it clearly verifies Eq.(\ref{detailed}) and shows the significance of $\int_{L_{t_0}}^{L_t} p_{\tau} \,  \D L_{\tau}$.

In Fig.\ref{fig:amp}  $A(\tau) =  \theta(t_0-\tau) \frac{\tau- t_0}{t-t_0} +1$ and  $B(\tau) = 1$. Contrary to the previous case there is a potential during the initial relaxation. This means  $t_0$ has to be chosen sufficiently big such that $\Pl{x,t_0,\lambda_{t_0}} \approx P(x,t,\lambda_{t})$. 
Fig.\ref{fig:amp}(a) shows that for the particularly chosen parameters  $t_0 = 10$ suffices. As before Fig.\ref{fig:amp}(b) verifies Eq.(\ref{detailed}) however 
$\int_{L_{t_0}}^{L_t} p_{\tau} \,  \D L_{\tau}$ seems to be negligible.

In Fig.\ref{fig:mov}  $A(\tau) =  5$ and  $B(\tau) = 5\, \theta(\tau - t_0) \frac{\tau- t_0}{t-t_0}$, again there is a potential during the initial relaxation
which we chose to be 5 $\kT$ deep. Instead of changing the amplitude $A(\tau)$ we are now changing the location of the potential. 
As in the previous case Fig.\ref{fig:mov}(a) shows good agreement of $\Pl{x,t_0,\lambda_{t_0}}$ with $P(x,t,\lambda_{t})$ and Fig.\ref{fig:mov}(b)
verifies Eq.(\ref{detailed}). However also in this  case
$\int_{L_{t_0}}^{L_t} p_{\tau} \,  \D L_{\tau}$ seems to be negligible.

In order to verify Eq. (\ref{qusteq}) and inequality (\ref{in}) we investigate the same examples as the quasi static limit is approached,
simply by making the duration of the protocol successively larger.
The results depicted in Fig. (\ref{fig:quasistat_sin}-\ref{fig:quasistat_mov}) are in good agreement with Eq. (\ref{qusteq}) and inequality (\ref{in}).
Note that in Fig. (\ref{fig:quasistat_sin}-\ref{fig:quasistat_mov}) the initial relaxation time $t_0$  was chosen large enough in order to reduce the error from approximating the initial distribution with $\Pl{x,t_0}$.
It is also important to realize that in  Fig.\ref{fig:quasistat_sin}  and Fig.\ref{fig:quasistat_amp}   $|\ex{W}-\int_{L_{t_0}}^{L_t} p_{\tau} \,  \D L_{\tau}-\Delta G |$ goes to zero faster than $\ex{W}$.

Interestingly the average work for the sinusoidal changing $A(\tau)$ is always negative even though the change of the potential is cyclic, see Fig.\ref{fig:quasistat_sin}.
As already discussed in the previous section this is not possible for confined systems since it would violate the second law of thermodynamics. However, for our non confined systems this is per se not a violation of the second law since
the system does not return to its original state. Repeating a cycle 
$n$ times does not necessarily lead to an infinite energy output, it depends on how the average work per cycle $\ex{W}_n$ behaves with $n$. And indeed as one can see from Fig.\ref{fig:cycle}(a,b), $\ex{W}_n$
 increases exponentially fast and decays to zero from above after a small but positive value has been reached.  This behavior leads to a positive total work $\ex{W}=\sum_{n}\ex{W}_n > 0$, for large enough $n$, see Fig. \ref{fig:cycle}(c).

\section{Conclusion and Discussion}
We have derived a work fluctuation theorem, see Eq. (\ref{work_int}), similar to the Jarzynski equality but applicable to a Brownian particle inside a potential well with finite depth that is changed in time by an external protocol.
Such systems are not able to reach thermal equilibrium which is reflected in the fluctuation theorem by an additional path dependent term (\ref{new_term})  besides work. The inequality that results from this fluctuation theorem puts a fundamental lower bound on the work that is needed to change the potential in time. It  is expected to become an equality in the quasi-static limit which gives the new term the meaning of an energy that can be extracted from the never ending diffusive spreading of the system. 

The only approximation in the derivation of Eq. (\ref{work_int}) is done by approximating the PDF at the start of the protocol with the long time asymptotic density $\Pl{x,t_0,\lambda_{t_0}}$ given by Eq.\eqref{uniformdens}.
Our  theory would be exact, if the density at the beginning of the
protocol were exactly the Gauss-Boltzmann density $\Pl{x,t_0,\lambda_{t_0}}$. This approximation is the better the longer the initial time evolution. So for every finite time evolution $t_0$ also relation  (\ref{work_int}) is only an approximation.
At first glance this seems to be a disadvantage in comparison to the Jarzynski equality. Here the Boltzmann density, which is an exact solution of the Fokker-Planck equation, is assumed to describe the system at the start of the protocol. However this line of thought is misleading. In Brownian dynamics simulations or an experiment one would need to wait infinitely long for a confined system to reach a state which is exactly described by the Boltzmann density. In that sense assuming that a confined system can be described by the Boltzmann density is as much of an approximation as assuming that a non-confined system can be described by $\Pl{x,t_0,\lambda_{t_0}}$. The rate of convergence however might be different.

A major open question is how Eq. (\ref{work_int}) relates to stochastic thermodynamics and one of its main results, the Seifert fluctuation theorem  \cite{seifert2005entropy}.
Considering the simple special case of free Brownian motion it is easy to show that they do not coincide. Furthermore the inequality implied by Seifert's theorem becomes 
an equality if the system is time reversible, inequality (\ref{in}) on the other hand is expected to become an equality if the protocol is quasi-static. 
Now, for stochastic systems the Jarzynski equality can be seen as a special case of Seifert's fluctuation theorem. Its implied inequality becomes an equality in the quasi-static limit which in this case is also the time reversible limit. The conclusion here would be that for non-confined systems time reversibility is no longer implied by quasi-staticity. Intuitively this is simply a consequence of the never ending diffusive spreading. However, in order to make a more definite statement further investigations are required. 

Another question is when the term \eqref{new_term} in the fluctuation theorem \eqref{work_int} becomes irrelevant? It does not appear in the special case of the infinitely fast protocol, see
Eq. \eqref{flucttheo}. It also seems to be irrelevant in the numerical examples where the initial relaxation time is much longer than the duration of the protocol, see Fig.\ref{fig:amp}(b) and Fig.\ref{fig:mov}(b). Both of these results point in the direction that \eqref{new_term} is negligible if the  initial relaxation time is long compared to the duration of the protocol. 

Yet another question is how general these type of work fluctuation theorems are. In principle, the mathematical procedure based on the Feynman-Kac formula can be applied to 
any long time asymptotic initial PDF. Consequently the difficult part in deriving such a fluctuation theorem is to find this PDF.
Some already existing and usable results for further research are presented in \cite{defaveri2020regularized, dechant2011solution}.
\section{Acknowledgement}
We are grateful to E.Aghion,  E. Barkai, S. Bo and E. Lutz for valuable discussions.
\bibliography{infinitst.bib}

\begin{thebibliography}{23}%
\makeatletter
\providecommand \@ifxundefined [1]{%
 \@ifx{#1\undefined}
}%
\providecommand \@ifnum [1]{%
 \ifnum #1\expandafter \@firstoftwo
 \else \expandafter \@secondoftwo
 \fi
}%
\providecommand \@ifx [1]{%
 \ifx #1\expandafter \@firstoftwo
 \else \expandafter \@secondoftwo
 \fi
}%
\providecommand \natexlab [1]{#1}%
\providecommand \enquote  [1]{``#1''}%
\providecommand \bibnamefont  [1]{#1}%
\providecommand \bibfnamefont [1]{#1}%
\providecommand \citenamefont [1]{#1}%
\providecommand \href@noop [0]{\@secondoftwo}%
\providecommand \href [0]{\begingroup \@sanitize@url \@href}%
\providecommand \@href[1]{\@@startlink{#1}\@@href}%
\providecommand \@@href[1]{\endgroup#1\@@endlink}%
\providecommand \@sanitize@url [0]{\catcode `\\12\catcode `\$12\catcode
  `\&12\catcode `\#12\catcode `\^12\catcode `\_12\catcode `\%12\relax}%
\providecommand \@@startlink[1]{}%
\providecommand \@@endlink[0]{}%
\providecommand \url  [0]{\begingroup\@sanitize@url \@url }%
\providecommand \@url [1]{\endgroup\@href {#1}{\urlprefix }}%
\providecommand \urlprefix  [0]{URL }%
\providecommand \Eprint [0]{\href }%
\providecommand \doibase [0]{http://dx.doi.org/}%
\providecommand \selectlanguage [0]{\@gobble}%
\providecommand \bibinfo  [0]{\@secondoftwo}%
\providecommand \bibfield  [0]{\@secondoftwo}%
\providecommand \translation [1]{[#1]}%
\providecommand \BibitemOpen [0]{}%
\providecommand \bibitemStop [0]{}%
\providecommand \bibitemNoStop [0]{.\EOS\space}%
\providecommand \EOS [0]{\spacefactor3000\relax}%
\providecommand \BibitemShut  [1]{\csname bibitem#1\endcsname}%
\let\auto@bib@innerbib\@empty
%</preamble>
\bibitem [{\citenamefont {Jarzynski}(1997)}]{jarzynski1997nonequilibrium}%
  \BibitemOpen
  \bibfield  {author} {\bibinfo {author} {\bibfnamefont {C.}~\bibnamefont
  {Jarzynski}},\ }\href@noop {} {\bibfield  {journal} {\bibinfo  {journal}
  {Physical Review Letters}\ }\textbf {\bibinfo {volume} {78}},\ \bibinfo
  {pages} {2690} (\bibinfo {year} {1997})}\BibitemShut {NoStop}%
\bibitem [{\citenamefont {Seifert}(2012)}]{seifert2012stochastic}%
  \BibitemOpen
  \bibfield  {author} {\bibinfo {author} {\bibfnamefont {U.}~\bibnamefont
  {Seifert}},\ }\href@noop {} {\bibfield  {journal} {\bibinfo  {journal}
  {Reports on progress in physics}\ }\textbf {\bibinfo {volume} {75}},\
  \bibinfo {pages} {126001} (\bibinfo {year} {2012})}\BibitemShut {NoStop}%
\bibitem [{\citenamefont {Seifert}(2005)}]{seifert2005entropy}%
  \BibitemOpen
  \bibfield  {author} {\bibinfo {author} {\bibfnamefont {U.}~\bibnamefont
  {Seifert}},\ }\href@noop {} {\bibfield  {journal} {\bibinfo  {journal}
  {Physical review letters}\ }\textbf {\bibinfo {volume} {95}},\ \bibinfo
  {pages} {040602} (\bibinfo {year} {2005})}\BibitemShut {NoStop}%
\bibitem [{\citenamefont {Sagawa}\ and\ \citenamefont
  {Ueda}(2010)}]{sagawa2010generalized}%
  \BibitemOpen
  \bibfield  {author} {\bibinfo {author} {\bibfnamefont {T.}~\bibnamefont
  {Sagawa}}\ and\ \bibinfo {author} {\bibfnamefont {M.}~\bibnamefont {Ueda}},\
  }\href@noop {} {\bibfield  {journal} {\bibinfo  {journal} {Physical review
  letters}\ }\textbf {\bibinfo {volume} {104}},\ \bibinfo {pages} {090602}
  (\bibinfo {year} {2010})}\BibitemShut {NoStop}%
\bibitem [{\citenamefont {Hatano}\ and\ \citenamefont
  {Sasa}(2001)}]{hatano2001steady}%
  \BibitemOpen
  \bibfield  {author} {\bibinfo {author} {\bibfnamefont {T.}~\bibnamefont
  {Hatano}}\ and\ \bibinfo {author} {\bibfnamefont {S.-i.}\ \bibnamefont
  {Sasa}},\ }\href@noop {} {\bibfield  {journal} {\bibinfo  {journal} {Physical
  review letters}\ }\textbf {\bibinfo {volume} {86}},\ \bibinfo {pages} {3463}
  (\bibinfo {year} {2001})}\BibitemShut {NoStop}%
\bibitem [{\citenamefont {Crooks}(1999)}]{crooks1999entropy}%
  \BibitemOpen
  \bibfield  {author} {\bibinfo {author} {\bibfnamefont {G.~E.}\ \bibnamefont
  {Crooks}},\ }\href@noop {} {\bibfield  {journal} {\bibinfo  {journal}
  {Physical Review E}\ }\textbf {\bibinfo {volume} {60}},\ \bibinfo {pages}
  {2721} (\bibinfo {year} {1999})}\BibitemShut {NoStop}%
\bibitem [{\citenamefont {Esposito}\ and\ \citenamefont {Van~den
  Broeck}(2010)}]{esposito2010three}%
  \BibitemOpen
  \bibfield  {author} {\bibinfo {author} {\bibfnamefont {M.}~\bibnamefont
  {Esposito}}\ and\ \bibinfo {author} {\bibfnamefont {C.}~\bibnamefont {Van~den
  Broeck}},\ }\href@noop {} {\bibfield  {journal} {\bibinfo  {journal}
  {Physical review letters}\ }\textbf {\bibinfo {volume} {104}},\ \bibinfo
  {pages} {090601} (\bibinfo {year} {2010})}\BibitemShut {NoStop}%
\bibitem [{\citenamefont {Dabelow}\ \emph {et~al.}(2019)\citenamefont
  {Dabelow}, \citenamefont {Bo},\ and\ \citenamefont
  {Eichhorn}}]{dabelow2019irreversibility}%
  \BibitemOpen
  \bibfield  {author} {\bibinfo {author} {\bibfnamefont {L.}~\bibnamefont
  {Dabelow}}, \bibinfo {author} {\bibfnamefont {S.}~\bibnamefont {Bo}}, \ and\
  \bibinfo {author} {\bibfnamefont {R.}~\bibnamefont {Eichhorn}},\ }\href@noop
  {} {\bibfield  {journal} {\bibinfo  {journal} {Physical Review X}\ }\textbf
  {\bibinfo {volume} {9}},\ \bibinfo {pages} {021009} (\bibinfo {year}
  {2019})}\BibitemShut {NoStop}%
\bibitem [{\citenamefont {Jarzynski}(2011)}]{jarzynski2011equalities}%
  \BibitemOpen
  \bibfield  {author} {\bibinfo {author} {\bibfnamefont {C.}~\bibnamefont
  {Jarzynski}},\ }\href@noop {} {\bibfield  {journal} {\bibinfo  {journal}
  {Annu. Rev. Condens. Matter Phys.}\ }\textbf {\bibinfo {volume} {2}},\
  \bibinfo {pages} {329} (\bibinfo {year} {2011})}\BibitemShut {NoStop}%
\bibitem [{\citenamefont {Neri}(2020)}]{neri2020second}%
  \BibitemOpen
  \bibfield  {author} {\bibinfo {author} {\bibfnamefont {I.}~\bibnamefont
  {Neri}},\ }\href@noop {} {\bibfield  {journal} {\bibinfo  {journal} {Physical
  Review Letters}\ }\textbf {\bibinfo {volume} {124}},\ \bibinfo {pages}
  {040601} (\bibinfo {year} {2020})}\BibitemShut {NoStop}%
\bibitem [{\citenamefont {Neri}\ \emph {et~al.}(2019)\citenamefont {Neri},
  \citenamefont {Rold{\'a}n}, \citenamefont {Pigolotti},\ and\ \citenamefont
  {J{\"u}licher}}]{neri2019integral}%
  \BibitemOpen
  \bibfield  {author} {\bibinfo {author} {\bibfnamefont {I.}~\bibnamefont
  {Neri}}, \bibinfo {author} {\bibfnamefont {{\'E}.}~\bibnamefont
  {Rold{\'a}n}}, \bibinfo {author} {\bibfnamefont {S.}~\bibnamefont
  {Pigolotti}}, \ and\ \bibinfo {author} {\bibfnamefont {F.}~\bibnamefont
  {J{\"u}licher}},\ }\href@noop {} {\bibfield  {journal} {\bibinfo  {journal}
  {Journal of Statistical Mechanics: Theory and Experiment}\ }\textbf {\bibinfo
  {volume} {2019}},\ \bibinfo {pages} {104006} (\bibinfo {year}
  {2019})}\BibitemShut {NoStop}%
\bibitem [{\citenamefont {Aghion}\ \emph {et~al.}(2019)\citenamefont {Aghion},
  \citenamefont {Kessler},\ and\ \citenamefont {Barkai}}]{aghion2019non}%
  \BibitemOpen
  \bibfield  {author} {\bibinfo {author} {\bibfnamefont {E.}~\bibnamefont
  {Aghion}}, \bibinfo {author} {\bibfnamefont {D.~A.}\ \bibnamefont {Kessler}},
  \ and\ \bibinfo {author} {\bibfnamefont {E.}~\bibnamefont {Barkai}},\
  }\href@noop {} {\bibfield  {journal} {\bibinfo  {journal} {Physical review
  letters}\ }\textbf {\bibinfo {volume} {122}},\ \bibinfo {pages} {010601}
  (\bibinfo {year} {2019})}\BibitemShut {NoStop}%
\bibitem [{\citenamefont {Aghion}\ \emph {et~al.}(2020)\citenamefont {Aghion},
  \citenamefont {Kessler},\ and\ \citenamefont {Barkai}}]{aghion2020infinite}%
  \BibitemOpen
  \bibfield  {author} {\bibinfo {author} {\bibfnamefont {E.}~\bibnamefont
  {Aghion}}, \bibinfo {author} {\bibfnamefont {D.~A.}\ \bibnamefont {Kessler}},
  \ and\ \bibinfo {author} {\bibfnamefont {E.}~\bibnamefont {Barkai}},\
  }\href@noop {} {\bibfield  {journal} {\bibinfo  {journal} {Chaos, Solitons \&
  Fractals}\ }\textbf {\bibinfo {volume} {138}},\ \bibinfo {pages} {109890}
  (\bibinfo {year} {2020})}\BibitemShut {NoStop}%
\bibitem [{\citenamefont {Dechant}\ \emph {et~al.}(2011)\citenamefont
  {Dechant}, \citenamefont {Lutz}, \citenamefont {Barkai},\ and\ \citenamefont
  {Kessler}}]{dechant2011solution}%
  \BibitemOpen
  \bibfield  {author} {\bibinfo {author} {\bibfnamefont {A.}~\bibnamefont
  {Dechant}}, \bibinfo {author} {\bibfnamefont {E.}~\bibnamefont {Lutz}},
  \bibinfo {author} {\bibfnamefont {E.}~\bibnamefont {Barkai}}, \ and\ \bibinfo
  {author} {\bibfnamefont {D.}~\bibnamefont {Kessler}},\ }\href@noop {}
  {\bibfield  {journal} {\bibinfo  {journal} {Journal of Statistical Physics}\
  }\textbf {\bibinfo {volume} {145}},\ \bibinfo {pages} {1524} (\bibinfo {year}
  {2011})}\BibitemShut {NoStop}%
\bibitem [{\citenamefont {Wang}\ \emph {et~al.}(2019)\citenamefont {Wang},
  \citenamefont {Deng},\ and\ \citenamefont {Chen}}]{wang2019ergodic}%
  \BibitemOpen
  \bibfield  {author} {\bibinfo {author} {\bibfnamefont {X.}~\bibnamefont
  {Wang}}, \bibinfo {author} {\bibfnamefont {W.}~\bibnamefont {Deng}}, \ and\
  \bibinfo {author} {\bibfnamefont {Y.}~\bibnamefont {Chen}},\ }\href@noop {}
  {\bibfield  {journal} {\bibinfo  {journal} {The Journal of chemical physics}\
  }\textbf {\bibinfo {volume} {150}},\ \bibinfo {pages} {164121} (\bibinfo
  {year} {2019})}\BibitemShut {NoStop}%
\bibitem [{\citenamefont {Aaronson}(1997)}]{aaronson1997introduction}%
  \BibitemOpen
  \bibfield  {author} {\bibinfo {author} {\bibfnamefont {J.}~\bibnamefont
  {Aaronson}},\ }\href@noop {} {\emph {\bibinfo {title} {An introduction to
  infinite ergodic theory}}},\ \bibinfo {number} {50}\ (\bibinfo  {publisher}
  {American Mathematical Soc.},\ \bibinfo {year} {1997})\BibitemShut {NoStop}%
\bibitem [{\citenamefont {Leibovich}\ and\ \citenamefont
  {Barkai}(2019)}]{leibovich2019infinite}%
  \BibitemOpen
  \bibfield  {author} {\bibinfo {author} {\bibfnamefont {N.}~\bibnamefont
  {Leibovich}}\ and\ \bibinfo {author} {\bibfnamefont {E.}~\bibnamefont
  {Barkai}},\ }\href@noop {} {\bibfield  {journal} {\bibinfo  {journal}
  {Physical Review E}\ }\textbf {\bibinfo {volume} {99}},\ \bibinfo {pages}
  {042138} (\bibinfo {year} {2019})}\BibitemShut {NoStop}%
\bibitem [{\citenamefont {Meyer}\ and\ \citenamefont
  {Kantz}(2017)}]{meyer2017infinite}%
  \BibitemOpen
  \bibfield  {author} {\bibinfo {author} {\bibfnamefont {P.}~\bibnamefont
  {Meyer}}\ and\ \bibinfo {author} {\bibfnamefont {H.}~\bibnamefont {Kantz}},\
  }\href@noop {} {\bibfield  {journal} {\bibinfo  {journal} {Physical Review
  E}\ }\textbf {\bibinfo {volume} {96}},\ \bibinfo {pages} {022217} (\bibinfo
  {year} {2017})}\BibitemShut {NoStop}%
\bibitem [{\citenamefont {Akimoto}\ \emph {et~al.}(2020)\citenamefont
  {Akimoto}, \citenamefont {Barkai},\ and\ \citenamefont
  {Radons}}]{akimoto2020infinite}%
  \BibitemOpen
  \bibfield  {author} {\bibinfo {author} {\bibfnamefont {T.}~\bibnamefont
  {Akimoto}}, \bibinfo {author} {\bibfnamefont {E.}~\bibnamefont {Barkai}}, \
  and\ \bibinfo {author} {\bibfnamefont {G.}~\bibnamefont {Radons}},\
  }\href@noop {} {\bibfield  {journal} {\bibinfo  {journal} {Physical Review
  E}\ }\textbf {\bibinfo {volume} {101}},\ \bibinfo {pages} {052112} (\bibinfo
  {year} {2020})}\BibitemShut {NoStop}%
\bibitem [{\citenamefont {Hummer}\ and\ \citenamefont
  {Szabo}(2001)}]{hummer2001free}%
  \BibitemOpen
  \bibfield  {author} {\bibinfo {author} {\bibfnamefont {G.}~\bibnamefont
  {Hummer}}\ and\ \bibinfo {author} {\bibfnamefont {A.}~\bibnamefont {Szabo}},\
  }\href@noop {} {\bibfield  {journal} {\bibinfo  {journal} {Proceedings of the
  National Academy of Sciences}\ }\textbf {\bibinfo {volume} {98}},\ \bibinfo
  {pages} {3658} (\bibinfo {year} {2001})}\BibitemShut {NoStop}%
\bibitem [{\citenamefont {Boksenbojm}\ \emph {et~al.}(2010)\citenamefont
  {Boksenbojm}, \citenamefont {Wynants},\ and\ \citenamefont
  {Jarzynski}}]{boksenbojm2010nonequilibrium}%
  \BibitemOpen
  \bibfield  {author} {\bibinfo {author} {\bibfnamefont {E.}~\bibnamefont
  {Boksenbojm}}, \bibinfo {author} {\bibfnamefont {B.}~\bibnamefont {Wynants}},
  \ and\ \bibinfo {author} {\bibfnamefont {C.}~\bibnamefont {Jarzynski}},\
  }\href@noop {} {\bibfield  {journal} {\bibinfo  {journal} {Physica A:
  Statistical Mechanics and its Applications}\ }\textbf {\bibinfo {volume}
  {389}},\ \bibinfo {pages} {4406} (\bibinfo {year} {2010})}\BibitemShut
  {NoStop}%
\bibitem [{\citenamefont {Brady}(1993)}]{brady1993brownian}%
  \BibitemOpen
  \bibfield  {author} {\bibinfo {author} {\bibfnamefont {J.~F.}\ \bibnamefont
  {Brady}},\ }\href@noop {} {\bibfield  {journal} {\bibinfo  {journal} {The
  Journal of chemical physics}\ }\textbf {\bibinfo {volume} {98}},\ \bibinfo
  {pages} {3335} (\bibinfo {year} {1993})}\BibitemShut {NoStop}%
\bibitem [{\citenamefont {Defaveri}\ \emph {et~al.}(2020)\citenamefont
  {Defaveri}, \citenamefont {Anteneodo}, \citenamefont {Kessler},\ and\
  \citenamefont {Barkai}}]{defaveri2020regularized}%
  \BibitemOpen
  \bibfield  {author} {\bibinfo {author} {\bibfnamefont {L.}~\bibnamefont
  {Defaveri}}, \bibinfo {author} {\bibfnamefont {C.}~\bibnamefont {Anteneodo}},
  \bibinfo {author} {\bibfnamefont {D.~A.}\ \bibnamefont {Kessler}}, \ and\
  \bibinfo {author} {\bibfnamefont {E.}~\bibnamefont {Barkai}},\ }\href@noop {}
  {\bibfield  {journal} {\bibinfo  {journal} {arXiv preprint arXiv:2004.04325}\
  } (\bibinfo {year} {2020})}\BibitemShut {NoStop}%
\end{thebibliography}%

\end{document}